\documentclass[12pt]{article}
\usepackage{amsmath,amsfonts}
\makeatletter \@addtoreset{equation}{section}

\usepackage{cite}
\usepackage{bm}
\usepackage{dcolumn}
\makeatother
\def\one{{\hbox{ 1\kern-.8mm l}}}
\newcommand{\Dslash}{\not{\hbox{\kern-4pt $D$}}}
\newcommand{\pdslash}{\not{\hbox{\kern-2pt $\partial$}}}
\newcommand{\be}{\begin{equation}}
\newcommand{\bea}{\begin{eqnarray}}
\newcommand{\eea}{\end{eqnarray}}
\newcommand{\ba}{\begin{array}}
\newcommand{\ea}{\end{array}}
\newcommand{\ee}{\end{equation}}
\newcommand{\nn}{\nonumber}

\begin{document}

\begin{titlepage}
\vspace*{1mm}%
\hfill%
\vbox{
    \halign{#\hfil        \cr
           IPM/P-2009/nnn \cr
                     } 
      }  
\vspace*{15mm}%
\begin{center}

{{\Large {\bf On three dimensional bigravity }}}

\vspace*{15mm} \vspace*{1mm} {Hamid R. Afshar$^{a,b}$, Mohsen Alishahiha$^{a}$ and Ali Naseh$^{a,b}$}

 \vspace*{1cm}

{\it ${}^a$ School of physics, Institute for Research in Fundamental Sciences (IPM)\\
P.O. Box 19395-5531, Tehran, Iran \\ }

\vspace*{.4cm}

{\it ${}^b$ Department of Physics, Sharif University of Technology \\
P.O. Box 11365-9161, Tehran, Iran}

\vspace*{2cm}
\end{center}

\begin{abstract}
In this paper we explore some features of $f$-$g$ theory in three dimensions.
We show that the theory has (A)dS and (A)dS wave solutions. In particular at a critical value of the
coupling constant we see that the model admits Log gravity solution as well, reminiscing TMG and NMG.
We have also studied a class of exact static spherically symmetric black hole
solution in the model.

\end{abstract}

\end{titlepage}


\section{Introduction}

Although in three dimensional pure gravity there is no dynamical propagating mode, in the presence of a
negative cosmological constant the theory admits rather a non-trivial solution known as BTZ black hole.
Therefore quantum mechanically the content of the theory must be quite non-trivial in order to be able to
describe the entropy of the black hole solutions. Actually there are lots of activities to understand
quantum gravity in three dimensions, though it is not quite clear what that really means.

Since any solution in three dimensional pure gravity in the presence of negative cosmological constant
is locally $AdS_3$ one may wonder that the quantum gravity in three dimensions may be defined
by a two dimensional CFT theory via AdS/CFT correspondence\cite{Maldacena:1997re,Brown:1986nw}.
Indeed the first attempt toward this suggestion was made
by Witten\cite{Witten:2007kt}, though due to a non-trivial assumption of the holomorphic
factorization it is not clear whether
the suggestion could directly be applied to pure gravity in three dimensions.

It is, however, possible to have a chiral gravity in which the holomorphic factorization could be a natural
property of the theory. Indeed the three dimensional chiral gravity has first been introduced in \cite{Li:2008dq}
 in the context of TMG model \cite{{Deser:1981wh},{Deser:1982vy}} where the gravitational
Chern-Simons term has added to Einstein gravity. It was shown \cite{Li:2008dq} that to have a consistent stable theory above
the AdS vacuum solution the coupling constant of the Chern-Simons term has to be tuned to a critical value
leading to a chiral theory ( See for example \cite{{Carlip:2008jk},{Compere:2008us},{Grumiller:2008qz},
{Sachs:2008gt},{Carlip:2008eq},
{Giribet:2008bw},{Grumiller:2008es},{Alishahiha:2008rt},{Henneaux:2009pw},{Maloney:2009ck}}
 for different discussions and aspects of chiral gravity.). TMG has also several novel solutions which
might eventually help us understanding the three dimensional gravity better.

In an attempt to explore three dimensional gravity a new model of massive gravity, known as NMG, has also been
introduced in \cite{Bergshoeff:2009hq} where the authors have shown that the theory enjoys very
similarity to TMG model but still has its own new features.

More recently, following the idea of the four dimensional $f$-$g$ theory \cite{Isham:1971gm},
even a new three dimensional
gravity has been introduced by Banados and Theisen \cite{Banados:2009it}. The model could be thought of as
three dimensional gravity coupled to a rank two tensor with a given interaction. The corresponding action
is
\cite{Banados:2009it}
\bea\label{action}
I[g_{\mu \nu},f_{\mu \nu}]=\frac{1}{16\pi G}\int d^3x\left[ \sqrt{-g} \left( R^g+\frac{2}{\ell^2}\right)
+\sigma\sqrt{-f}\left( R^f+\frac{2}{\ell^2}\right) - U_0(g,f) \right],
\eea
where the potential is given by
\bea
U_0(g,f)=\frac{\nu}{\ell^2}\sqrt{-f}\;(g_{\mu \nu}-f_{\mu \nu})(g_{\alpha \beta}-f_{\alpha \beta})
(f^{\mu\alpha}f^{\nu\beta}-f^{\mu\nu}f^{\alpha\beta}).
\eea
Here $\sigma, \nu$ and $\ell$ are constants that parameterize the theory. Due to special form of the
potential the theory has an $AdS_3$ vacuum solution given by $f_{\mu\nu}=g_{\mu\nu}=g_{\mu\nu}^{AdS}$.
It was shown in \cite{Banados:2009it} that upon expanding the action around this vacuum the theory at leading
order reduces to an NMG theory plus a trivial mode. Therefore one would expect to see some similarity
between the present theory and NMG theory. Different solutions of the model have also been studied in
\cite{Banados:2009it} where it was shown that the model admits another $AdS_3$ solution. In this new
AdS solution although $f$ and $g$ metrics are not equal, they are proportional to an $AdS_3$ metric.

In this article we would like to explore some features of $f$-$g$ gravity in three dimensions.
To proceed we will first assume that the $g$ and $f$ modes have different {\it bare cosmological constants}
denoted by $\Lambda_0^g/2$ and $\Lambda_0^f/2$, respectively.  Moreover following \cite{Banados:2009it} we will
consider a potential with a generalized measure as follows
\bea
U_m(g,f)=\frac{\nu}{\ell^2}(-g)^m(-f)^n(g^{\mu \nu}-f^{\mu \nu})(g^{\alpha \beta}-f^{\alpha \beta})
(f_{\mu\alpha}f_{\nu\beta}-f_{\mu\nu}f_{\alpha\beta})
\eea
with $m+n=\frac{1}{2}$. Here $\ell^2$ is a dimensionful parameter which to
avoid having more dimensionful parameter we assume that
it could be given in terms of $\Lambda_0^g$ and $\Lambda_0^f$.
Therefore we will consider a generalization of the action \eqref{action} given by
\bea\label{gaction}
I[g_{\mu \nu},f_{\mu \nu}]=\frac{1}{16\pi G}\int d^3x\left[ \sqrt{-g} \left( R^g-\Lambda_0^g\right)
+\sigma\sqrt{-f}\left( R^f-\Lambda_0^f\right) - U_m(g,f) \right],
\eea
The corresponding equations of motion coming from this action are\footnote{Here we set $16\pi G=1$.}
\be\label{eom}
G^g_{\mu\nu}+\frac{1}{2}\Lambda_0^gg_{\mu\nu}=T^g_{\mu\nu},\;\;\;\;\;\;\;\;\;\;\;\;\;
G^f_{\mu\nu}+\frac{1}{2}\Lambda_0^ff_{\mu\nu}=\frac{1}{\sigma} T^f_{\mu\nu},
\ee
where the energy momentum tensors, $ T^g_{\mu\nu},T^f_{\mu\nu}$ are \cite{Isham:1977rj}
\bea\label{tens}
T^g_{\mu\nu}&=&\frac{\nu}{\ell^2}\left(\frac{f}{g}\right)^n\bigg[m g_{\mu\nu}(g^{\alpha\beta}-
f^{\alpha\beta})(g^{\sigma\tau}-f^{\sigma\tau})(f_{\sigma\alpha}f_{\beta\tau}-f_{\alpha\beta}f_{\sigma\tau})
\nn\\&&\;\;\;\;\;\;\;\;\;\;\;\;\;\;\;\;\;\;-2(g^{\alpha\beta}-f^{\alpha\beta})(f_{\alpha\mu}f_{\beta\nu}-
f_{\alpha\beta}f_{\mu\nu})\bigg],\\
T^f_{\mu\nu}&=&\frac{\nu}{\ell^2}\left(\frac{g}{f}\right)^m\bigg[2 (g^{\alpha\beta}-f^{\alpha\beta})
(f_{\alpha\mu}f_{\beta\nu}-
f_{\alpha\beta}f_{\mu\nu})\nn\\
&&\;\;\;\;\;\;\;\;\;\;\;\;\;\;\;\;\;\;+(g^{\alpha\beta}-f^{\alpha\beta})(g^{\sigma\tau}-f^{\sigma\tau})
(nf_{\mu\nu}f_{\alpha\sigma}
f_{\beta\tau}-nf_{\mu\nu}f_{\alpha\beta}f_{\sigma\tau}\nn\\
&&\;\;\;\;\;\;\;\;\;\;\;\;\;\;\;\;\;\;+2f_{\alpha\mu}f_{\sigma\nu}f_{\beta\tau}
-2f_{\alpha\mu}f_{\beta\nu}f_{\sigma\tau})\bigg].\nn
\eea

The aim of this paper is to find different solutions of the above equations of motion. Indeed we will see that
the model admits several novel solutions reminiscing TMG and NMG models. We will also explore different aspects
of the solutions.

The paper is organized as follows. In the next section we find the AdS wave solution  in the model where we show
that at a critical value of the coupling constant, $\sigma$, the theory admits log gravity solution as well.
In section three we study the most general static spherically symmetric black hole solution in the model.
The last section is devoted to discussions.


\section{AdS wave solution}

\subsection*{(A)dS vacuums}
Motivated by the four dimensional $f$-$g$ theory \cite{Blas:2005yk} we can start from the anstaz $f_{\mu\nu}=
\gamma g_{\mu\nu}$ by which the equations of motion \eqref{eom} can be recast to the following form
\bea\label{Asol}
G^g_{\mu\nu}=\frac{1}{2}\Lambda^g g_{\mu\nu},\;\;\;\;\;\;\;\;\;\;G^f_{\mu\nu}=\frac{1}{2}\gamma \Lambda^f g_{\mu\nu},
\eea
where
\bea\label{lam}
\Lambda^g&=& \frac{4\nu}{\ell^2}\gamma^{3n}(\gamma-1)(2\gamma-3m(\gamma-1))-\Lambda_0^g\nn,\\
\Lambda^f&=& -\frac{4\nu}{\sigma\ell^2}\gamma^{-3m}(\gamma-1)(2\gamma+3n(\gamma-1))-\Lambda_0^f.
\eea
To have a consistent solution one needs to set $\Lambda^g=\gamma \Lambda^f$ which leads to the following
algebraic equation for $\gamma$
\bea\label{equa}
\frac{4\nu}{\ell^2}(\gamma-1)\gamma^{3n}\left(\left(2\gamma+3n(\gamma-1)\right)(1+\frac{1}{\sigma\sqrt{\gamma}})
-\frac{3}{2}(\gamma-1)\right)=\left(\Lambda_0^g-\gamma \Lambda_0^f\right).
\eea
In principle one could solve this algebraic equation to find $\gamma$ in terms of the parameters of the
theory. The most general solution will be $AdS_3$/$dS_3$ solution with the
radius of $\ell_{eff}=\sqrt{2/|\Lambda^g|}$ depending on the sign of $\Lambda^g$.
For the particular case of $\Lambda_0^g=\Lambda_0^f=-2/\ell^2$ the above equation reduces to
\bea
(\gamma-1)\left[\bigg(2\gamma+3n(\gamma-1)\bigg)\bigg(1+\frac{1}{\sigma\sqrt{\gamma}}\bigg)
-\frac{3}{2}(\gamma-1)-\frac{1}{2\nu}\gamma^{-3n}\right]=0.
\eea
Of course $\gamma=1$ is one of the solution of the above equation which indeed corresponds to the $AdS_3$ vacuum
solution considered in \cite{Banados:2009it}. On the other hand for $\Lambda_0^g\neq\Lambda_0^f$ one could still
simplify the equation by imposing the condition $\Lambda_0^g=\gamma \Lambda_0^f$ by which
the equation \eqref{equa} reduces to
\bea
\bigg(2\gamma+3n(\gamma-1)\bigg)\bigg(1+\frac{1}{\sigma\sqrt{\gamma}}\bigg)
=\frac{3}{2}(\gamma-1).
\eea
An interesting feature of this solution is that $\gamma$ is independent of $\nu$.

In the case of $\Lambda^g_0=\Lambda_0^f$ and $m=0$ it was shown in \cite{Banados:2009it} that
small fluctuations around the AdS vacuum of $f=g$ at leading order lead to an NMG theory plus a trivial
decoupled mode. It is then natural to see whether such  behavior can be seen in a more general solution as well.
To see this, following the four dimensional case \cite{Blas:2007zz,Blas:2007zza}, we consider the following
perturbations around the vacuum solution  given by $f=\gamma g$
\bea
g_{\mu\nu}=\bar{g}_{\mu\nu}+h_{\mu\nu},\;\;\;\;\;\;\;\;
f_{\mu\nu}=\gamma\left(\bar{g}_{\mu\nu}+\rho_{\mu\nu}\right),
\eea
where $\bar{g}_{\mu\nu}$ is the metric of $AdS_3$ solution with the radius $\ell_{eff}$.
The action for fluctuations becomes
\bea\label{Lin}
I[h_{\mu\nu},\rho_{\mu\nu}]=\frac{1}{16\pi G}\int\sqrt{-\overline{g}}\left(h^{\mu\nu}(\mathcal{G}h)_{\mu\nu}
+\sqrt{\gamma}\sigma\rho^{\mu\nu}(\mathcal{G}\rho)_{\mu\nu}-\gamma^{-3m}\frac{\nu}{\ell_{eff}^2}
(h-\rho)\cdot(h-\rho)\right)
\eea
where $\mathcal{G}$ is the Pauli-Fierz operator \cite{Fierz:1939ix} on curved $AdS_3$ background\footnote
{For simplicity we set $\ell=\ell_{eff}$.}
\bea
h^{\mu\nu}(\mathcal{G}h)_{\mu\nu}&\equiv&-\frac{1}{4} h_{\nu\rho;\mu}h^{\nu\rho;\mu}+\frac{1}{2} h_{\mu\nu;\rho}
h^{\rho\nu;\mu}-\frac{1}{2} h_{;\mu}h^{\mu\nu;\nu}+\frac{1}{4} h_{;\mu}h^{;\mu}\nn\\
&&+\frac{1}{2\ell_{eff}^2}\left(h_{\mu\nu}h^{\mu\nu}-\frac{1}{2}h^2\right).
\eea
Here the inner product is defined by
\bea
h\cdot h\equiv M\;h_{\mu\nu}h^{\mu\nu}-N\;h^2,
\eea
with
\be
M=-\gamma(\gamma-2),\;\;\;\;\;\;\;\;\;N=-3mn(\gamma-1)^2+2(n-m)\gamma(\gamma-1)+\gamma^2.
\ee
To proceed let us define two new modes $h^{(0)}$ and $h^{(m)}$ in terms of the original fluctuations
$h_{\mu\nu}$ and $\rho_{\mu\nu}$ as follows
\bea
\rho= h^{(0)}- h^{(m)},\;\;\;\;\;\;\;\;\;\;\;\;h=h^{(0)}+\sqrt{\gamma}\sigma h^{(m)}.
\eea
In terms of these newly defined modes the action of the fluctuations reads
\bea
I&=&\frac{1+\sqrt{\gamma}\sigma}{16\pi G}\int\sqrt{-\bar{g}}\left(
{h^{(0)}}^{\mu\nu}(\mathcal{G}h^{(0)})_{\mu\nu}\right)\nn\\
&&+\frac{(1+\sqrt{\gamma}\sigma)\sqrt{\gamma}\sigma}{16\pi G}\int\sqrt{-\bar{g}}\bigg[
{h^{(m)}}^{\mu\nu}(\mathcal{G}h^{(m)})_{\mu\nu}-\frac{1}{4}{\cal M}^2
\left({h^{(m)}}^{\mu\nu}h^{(m)}_{\mu\nu}-\frac{N}{M}(h^{(m)})^2
\right)\bigg],\nn
\eea
where
\bea\label{mass}
{\cal M}^2=\gamma^{-3m}\frac{4\nu}{\ell_e^2}\frac{1+\sqrt{\gamma}\sigma}{\sqrt{\gamma}\sigma}M.
\eea
Therefore in terms of new variables the theory represents a massless mode, $h^{(0)}_{\mu\nu}$, which
decouples from a massive mode,  $h^{(m)}_{\mu\nu}$. We note, however, that the action of the massive mode
is not in the deserved Pauli-Fierz form and therefore it might not in general be a ghost free theory
\cite{ArkaniHamed:2002sp,Dvali:2008em}.
To make it a ghost-free theory one must recast the action into the deserved
Pauli-Fierz form which can be done by further imposing another condition on $\gamma$. For example it can be done by
assuming that $M=N$ which can be solved by the following values of $\gamma$
\bea
\gamma=1,\;\;\;\;\;\;\;\;\;\;\;\;\;\;\gamma=\frac{3m(2m-1)}{6+m(6m-11)}.
\eea
With this condition the resulting theory is a single NMG with a new Newton constant
$\frac{1}{G'}=\frac{(1+\sqrt{\gamma}\sigma)\sqrt{\gamma}\sigma}{G}$ and the mass is given by \eqref{mass}
as in \cite{Bergshoeff:2009hq}.

Note that the solution of $\gamma=1$ is indeed the case considered in \cite{Banados:2009it}
for $\Lambda_0^g=\Lambda_0^f=-2/\ell^2$. We observe that for the generalized
effective potential it is also possible to get NMG theory even in the case of
$\Lambda_0^g\neq\Lambda_0^f$.\footnote{Note that, cases $m=0$ and $m=3/4$ lead to the case $\gamma=1$ for
which we have $\Lambda_0^g=\Lambda_0^f=-2/\ell^2$.}

\subsection*{AdS waves}

Having had the AdS vacuum solution, it is interesting to study
AdS wave solution in our model\footnote{ AdS wave solutions in TMG have been first found in \cite{{AyonBeato:2004fq},
{AyonBeato:2005qq}}.} . Actually the wave solution for $f$-$g$ theory in 4-dimensions have
been studied in \cite{Aichelburg:1971yz}.
Indeed as we have observed the $f$-$g$ theory at the linearized level
coincides with an NMG. On the other hand the existence of AdS wave configurations
for NMG has been explored in \cite{AyonBeato:2009yq}. Therefore it would be natural to look for
an AdS wave solution in our model too.

To proceed we consider an anstaz for the AdS wave solution in terms of the AdS metric as follows
\bea\label{wave}
g_{\mu\nu}=\bar{g}_{\mu\nu}+Hk_\mu k_\nu,\;\;\;\;\;\;\;\;\;
f_{\mu\nu}=\gamma(\bar{g}_{\mu\nu}+Fk_\mu k_\nu).
\eea
Here $k^{\mu}$ is a null vector field with respect to the metric $\bar{g}_{\mu\nu}$ where
$\bar{g}_{\mu\nu}$ is the ${AdS}_3$ metric obeying
\be\label{lam2}
\bar{G}_{\mu\nu}=\frac{1}{2}\Lambda^g \bar{g}_{\mu\nu}.
\ee
Moreover  $F$, $H$ are arbitrary functions which are
independent of the integral parameter along $k^{\mu}$.
Taking into account that the $k_\mu$ is a null vector, the equations of motion \eqref{eom} can be decomposed into
the following form
\be\label{eomwave}
\bar{G}_{\mu\nu}+\mathcal{G}^H_{\mu\nu}=\frac{1}{2}\Lambda^g
\bar{g}_{\mu\nu}+T^H_{\mu\nu},\;\;\;\;\;\;\;\;\;
\bar{G}_{\mu\nu}+\mathcal{G}^F_{\mu\nu}=\frac{1}{2}
\gamma\Lambda^f\bar{g}_{\mu\nu}+\frac{1}{\sigma}T^F_{\mu\nu},
\ee
where
\bea\label{tens2}
T^H_{\mu\nu}&=& \frac{2\nu}{\ell^2}\gamma^{3n}\big[\gamma^2(F-H)-3m(\gamma-1)^2H
\big]k_{\mu}k_{\nu},\nn\\
T^F_{\mu\nu}&=& -\frac{2\nu}{\ell^2}\gamma^{-3m+1}\big[
\gamma(\gamma-2)(H+3F)+4F+3n(\gamma-1)^2F\big]k_{\mu}k_{\nu},\nn
\eea
and\footnote{Since $k_{\mu}$ is null, The exact form of $\mathcal{G}^H_{\mu\nu}$, is obtained from the
linearized form of Einstein equation by replacing $h_{\mu\nu}$ with $Hk_{\mu}k_{\nu}$. For the linearized form
 of Einstein equation see for example \cite{Deser:2003vh}.}
\bea\label{nul}
\mathcal{G}^H_{\mu\nu}&=&-\frac{1}{2}\bar{\nabla}^2(Hk_{\mu}k_{\nu})+\frac{1}{2}\bar{\nabla}^{\sigma}
\bar{\nabla}_{\nu}
(Hk_{\sigma}k_{\mu})+\frac{1}{2}\bar{\nabla}^{\sigma}\bar{\nabla}_{\mu}
(Hk_{\sigma}k_{\nu})\cr &&-\frac{1}{2}\bar{g}_{\mu\nu}\bar{\nabla}_{\sigma}\bar{\nabla}_{\rho}
(Hk^{\sigma}k^{\rho})-\Lambda_0^gHk_{\mu}k_{\nu},\\\nn
\mathcal{G}^F_{\mu\nu}&=&-\frac{1}{2}\bar{\nabla}^2(Fk_{\mu}k_{\nu})+\frac{1}{2}\bar{\nabla}^{\sigma}
\bar{\nabla}_{\nu}
(Fk_{\sigma}k_{\mu})+\frac{1}{2}\bar{\nabla}^{\sigma}\bar{\nabla}_{\mu}
(Fk_{\sigma}k_{\nu})\cr &&-\frac{1}{2}\bar{g}_{\mu\nu}\bar{\nabla}_{\sigma}\bar{\nabla}_{\rho}
(Fk^{\sigma}k^{\rho})-\Lambda_0^fFk_{\mu}k_{\nu}.\nn
\eea
$\Lambda^g$, $\Lambda^f$ are also given in the equation (\ref{lam}). From the equation (\ref{lam2}) one needs to impose
the condition $\Lambda_g=\gamma \Lambda_f$ which determines $\gamma$ as in the previous case. On the other hand
by making use of the equation
\eqref{lam2} one arrives at
\be\label{neom}
\mathcal{G}^H_{\mu\nu}=T^H_{\mu\nu},\;\;\;\;\;\;\;\;\;\;\;\;
\mathcal{G}^F_{\mu\nu}=\frac{1}{\sigma}T^F_{\mu\nu}
\ee
To go further one needs to use an explicit parametrization for AdS geometry. Therefore for the
rest of our calculations we use the AdS solution in the Poincar\'{e} coordinates whose metric is given by
\bea
ds_{\bar{g}}^2&=&\frac{\ell_{eff}^2}{y^2}(dy^2-2dudv)\nn
\eea
where $\ell_{eff}$ is the effective AdS radius defined previously in terms of $\Lambda^g$.
In this notation one has
\bea\label{Wsol}
ds_g^2&=&\frac{\ell_{eff}^2}{y^2}(dy^2-2dudv+H(u,y)du^2),\cr
ds_f^2&=&\frac{\gamma\ell_{eff}^2}{y^2}(dy^2-2dudv+F(u,y)du^2).
\eea
Using the explicit form of the metrics and, setting $\Lambda_0^g=-2/\ell_1^2,\;\Lambda_0^f=-2/\ell^2_2$,  the nonzero
components
of $\mathcal{G}^H_{\mu\nu}$ and $\mathcal{G}^H_{\mu\nu}$ are
\bea
-2\;\mathcal{G}^H_{uu}&=&H''-\frac{1}{y}H'+\frac{4}{y^2}\left(1-\frac{\ell_{eff}^2}{\ell_1^2}\right)H\nn\\
-2\;\mathcal{G}^F_{uu}&=&F''-\frac{1}{y}F'+\frac{4}{y^2}\left(1-\gamma\frac{\ell_{eff}^2}{\ell_2^2}\right)F.
\eea
Therefore we get  the following
differential equations for $F$ and $H$
\bea
y^2H''(u, y)-y H'(u, y)+4\left(1-\frac{\ell_{eff}^2}{\ell_1^2}\right)H
&=&4\nu\gamma^{3n}\big[3m(\gamma-1)^2H+\gamma^2(H-F)
\big],\nn\\
y^2F''(u, y)-y F'(u, y)+4\left(1-\gamma\frac{\ell_{eff}^2}{\ell_2^2}\right)F&=&\frac{4\nu}{\sigma}\gamma^{-3m+1}
\big[3n(\gamma-1)^2F+
\gamma(\gamma-2)(H+3F)\cr
&&\;\;\;\;\;\;\;\;\;\;\;\;\;\;\;\;\;+4F\big],
\eea
where prime denotes the derivative with respect to $y$.
It is then clear that the most general solution of the above differential equations are in the form of $y^{\alpha}$
with constant $\alpha$.

Although we may
solve the above differential equations for general $\ell_1$ and $\ell_2$, for simplicity we set $\gamma=1$ which
results to $\ell_1=\ell_2=\ell_{eff}$. In this case one has
\bea
y^2H''(u, y)-y H'(u, y)&=&4\nu(H(u, y)-F(u, y)),\nn\\
y^2F''(u, y)-y F'(u, y)&=&\frac{4\nu}{\sigma}(F(u, y)-H(u, y)).\nn
\eea
For $\nu\neq-\frac{\sigma}{4(\sigma+1)}$ the above equations are solved by
\bea
F(u, y) &=& f_1(u)+f_2(u)y^2+f_3(u)y^{{1+\sqrt{1+4\nu(1+1/\sigma)}}}
+f_4(u)y^{{1-\sqrt{1+4\nu(1+1/\sigma)}}},\\
H(u, y) &=& f_1(u)+f_2(u)y^2-\sigma f_3(u)y^{{1+\sqrt{1+4\nu(1+1/\sigma)}}}
-\sigma f_4(u)y^{{1-\sqrt{1+4\nu(1+1/\sigma)}}},
\eea
where $f_i(u)$'s are arbitrary functions of $u$. While at the critical value
\bea
\nu=-\frac{\sigma}{4(\sigma+1)},
\eea
we get {Log} gravity solution as follows
\bea
F(u, y) &=& f_1(u)+f_2(u)y^2+y\bigg(f_3(u)\ln(y)+f_4(u)\bigg),\\
H(u, y) &=& f_1(u)+f_2(u)y^2-\sigma y\bigg(f_3(u)\ln(y)+f_4(u)\bigg).
\eea

To conclude this section we note that the $f$-$g$ model we are considering has two distinctive
vacua; $AdS_3$ vacuum and $AdS_3$ wave vacuum. Therefore it is also expected to get black hole
solutions out of these vacua by an identification. This is very similar to that in TMG and NMG
\cite{{AyonBeato:2004fq},{AyonBeato:2005qq}, {AyonBeato:2009yq}}
where one gets AdS and AdS wave solutions as well as their corresponding black holes which can be
obtained by an identification. Even the subtlety of having Log gravity, appears in all models too.

We note, however, that although in TMG and NMG we have another vacuum
known as warped $AdS_3$ solution, in our model we have not been able to find it yet. Of course in
\cite{Banados:2009it} the authors have found a new black hole solution whose asymptotics has
 $SL(2,R)\times R$ isometry and
 might be thought of as warped solution. If one thinks of this solution as an identification of a
vacuum solution then one might wonder that the model has also warped AdS solutions as well. It would be
interesting to see if such a solution can be found in this model too.

\section{Black hole solution}

In this section we would like to study a class of exact static spherically symmetric black hole solutions
of the three dimensional $f$-$g$ model given by the action \eqref{gaction}. To solve the corresponding
equations of motion \eqref{eom} we may start from an anstaz with an appropriate symmetries. Indeed, following
\cite{Isham:1977rj}, by making use of a suitable choice of coordinates, the most general static spherically
symmetric ansatz for the $f$ and $g$ metrics can be written as
\bea\label{Bsol}
ds^2_f&=&-Jdt^2+Kdr^2+r^2d\varphi^2,\nn\\
ds^2_g&=&-Cdt^2+2Ddtdr+Adr^2+Bd\varphi^2.
\eea
Note that, since we are looking for spherically symmetric solutions, the unknown parameters
$A$, $B$, $C$, $D$, $J$, $K$  can only be functions of $r$.

It is then straightforward to plug this
ansatz to the equations of motion \eqref{eom} to find the unknown functions. Indeed using the $tt$ and $tr$
components of the equations of motion of the metric $f$ one gets
\bea
\left(\frac{r^2}{2B}-1\right)D=0,
\eea
which for $D\neq 0$\footnote{For $D=0$ a class of solutions, indeed, reduces to that considered in the previous
subsection.} can be used to find $B=r^2/2$.

On the other hand using the explicit form of the metrics
as well as the fact that $B=r^2/2$ one finds different relations between the components of the energy momentum
tensors as follows \cite{Isham:1977rj}
\bea
C^{-1}T^{f}_{tt}=-D^{-1}T^{f}_{tr}=-A^{-1}T^{f}_{rr},\;\;\;\;\;\;\;\;\;\;\;\;
J^{-1}T^{g}_{tt}=-K^{-1}T^{g}_{rr},
\eea
which together with the equations of motion can be used to find different relations between the
components of the Ricci tensors
\bea\label{Ricc}
AR^{f}_{tt}+CR^{f}_{rr}=0,\;\;\;\;\;\;\;\;\;\;\;\;\;\;\;\;
KR^{g}_{tt}+JR^{g}_{rr}=0.
\eea
Using the explicit form of the Ricci tensors of the anstaz \eqref{Bsol} one finds
\bea
(JK)'=0,\;\;\;\;\;\;\;\;\;\;\Delta'=0,\;\;\;\;\;{\rm with}\;\;\;\;\Delta=AC+D^2,
\eea
That means $\Delta$ and $JK$ are constants of motion. It is always possible to set
$KJ=1$ by a reparametrization of the time coordinate $t$. With this information, setting $\Lambda_0^g=-2/\ell_1^2$ and
$\Lambda_0^f=-2/\ell_2^2$, it is then easy
to solve the rest of the equations of motion to find
\bea\label{Bsol1}
ds_f^2=f_{\mu\nu}dx^{\mu}dx^{\nu}&=&-\left({\tilde \Lambda}_fr^2-M_f\right)dt^2
+\frac{dr^2}{\left({\tilde \Lambda}_fr^2-M_f\right)}+r^2d\varphi^2\nn\cr
ds_g^2=g_{\mu\nu}dx^{\mu}dx^{\nu}&=&-\Delta\left({\tilde \Lambda}_gr^2-M_g\right)
dt^2+\frac{Y\;dr^2}{\left({\tilde \Lambda}_fr^2-M_f\right)}
+\frac{r^2}{2}d\varphi^2\\ &&+2\sqrt{\Delta\left(1- XY\right)}dtdr\nn
\eea
with
\bea
X=\frac{{\tilde \Lambda}_gr^2-M_g}{{\tilde \Lambda}_fr^2-M_f},\;\;\;\;\;\;\;Y=\frac{1}{2}+\Delta(2-X).
\eea
Here
\bea
{\tilde \Lambda}_g&=&\frac{\nu}{\ell^2}\left(\frac{2}{\Delta}\right)^{1+n}\left(1+m(\Delta-1) \right)+\frac{1}
{\ell_1^2}\nn\\
{\tilde \Lambda}_f&=&
-\frac{\nu}{\sigma\ell^2}\left(\frac{2}{\Delta}\right)^{1-m}\left(1-n(\Delta-1)\right)+\frac{1}{\ell_2^2}
\eea
This is a black hole solution which is different from that obtained in \cite{Banados:2009it}.
Actually the metric $f$ represents a BTZ black hole whose horizon is located at
\bea
r_H^2=\frac{M_f}{{\tilde \Lambda}_f},
\eea
and its Hawking temperature is
\be
T_H=\frac{\sqrt{{\tilde \Lambda}_f M_f}}{2\pi}
\ee
It is worth noting that at the particular value of $X=2$ both metrics become diagonal and that of
BTZ black hole. This is indeed the BTZ black hole of the previous section. Note that in this case we have
\be
{\tilde \Lambda}_g=2{\tilde \Lambda}_f,\;\;\;\;\;\;\;\;\;M_g=2M_f.
\ee

It is also instructive to study the asymptotic behavior of the above black hole solution. In fact for
large $r$ the solution (\ref{Bsol1}) asymptotically approaches to
\bea\label{asy}
ds_f^2&\simeq&-{\tilde \Lambda}_fr^2dt^2+\frac{dr^2}{{\tilde \Lambda}_fr^2}+r^2d\varphi^2\nn\\
ds_g^2&\simeq&-\Delta{\tilde \Lambda}_gr^2dt^2+\frac{y}{{\tilde \Lambda}_f}\frac{dr^2}{r^2}+2\delta dt dr+
\frac{r^2}{2}d\varphi^2
\eea
where
\be
y=\frac{1}{2}+\Delta(2-\frac{{\tilde \Lambda}_g}{{\tilde \Lambda}_f}),\;\;\;\;\;\;\;\;\;\;\;\;\delta^2=
\Delta(1-\frac{{\tilde \Lambda}_g}{{\tilde \Lambda}_f}y).
\ee
In order to have asymptotically well behaved AdS solution one needs to impose the condition,
$\delta\rightarrow 0$ while we are taking the limit of $r\rightarrow \infty$.
In other words for large $r$ one should have $y\rightarrow \frac{{\tilde \Lambda}_f}{{\tilde \Lambda}_g}$.

It is worth noting that although for the case of $y=\frac{{\tilde \Lambda}_f}{{\tilde \Lambda}_g}$
both metrics have 6 Killing vectors, only four of them are common and therefore the whole solution has
$SL(2,R)\times R$ isometry. On the other hand for $y=\frac{1}{2}$ and $\Delta=\frac{1}{4}$ the isometry
enhances to $SL(2,R)\times SL(2,R)$.

As a result we note that the black hole solution we have found may be thought of as a descendant
of the AdS vacuum. It is then natural to look for other black hole solutions which could be related
to AdS wave as well as log gravity solutions of the previous section.

\section{Discussions}
In this paper we have considered $f$-$g$ gravity in three dimensions. We have found different solutions
of the equations of motion, including (A)dS, AdS wave and static spherically symmetric black hole solutions.
More precisely we have found two distinctive vacuum solutions for our model; AdS and AdS wave solutions.
This is very similar to what has been found  for
TMG and NMG models\cite{{AyonBeato:2004fq},{AyonBeato:2005qq}, {AyonBeato:2009yq}}.
Indeed, even the subtlety of having Log gravity
at a critical value of the coupling constant appears in our model too. We note, however, that in comparison
with TMG and NMG we have not been able to find the warped AdS vacuum solution in our model. Of course
due to the black hole solution whose asymptotics is warped in \cite{Banados:2009it}, one might expect to have
warped AdS vacuum in our model too.

A similarity with NMG might be understood from the
fact that small fluctuations around the AdS vacuums of the $f$-$g$ gravity at leading order lead to
an NMG mode plus a free decoupled mode.

In this paper we have not discussed about the conserved and asymptotic charges of the black hole
solution in $f$-$g$ model. We note, however, that it is an important issue specially if we would like
to find a CFT dual for $f$-$g$ gravity in three dimensions. Of course for AdS vacuums in section 2
we would expect that the  asymptotic symmetry
of these solutions would be Virasoro algebra with central charge $c_m=\frac{3\ell_{eff}}{2G}(1+\sigma\sqrt{\gamma})$,
which reduces to that in \cite{Banados:2009it} for $\gamma=1$.

Following \cite{Banados:2009it} the $f$-$g$ model we have considered was based on the fact that
the action of both metrics $f$ and $g$ are given by Hilbert-Einstein gravity. Of course the model is equipped
with an effective potential which represents the interaction between $f$ and $g$.
We note, however, that in three dimensions there is another natural action one may consider for the
gravity; namely the gravitational Chern-Simons action. Therefore we may use the Chern-Simons action
for one or both of the metrics. To be specific let us consider an $f$-$g$ gravity in which the action of
$g$ metric is given by Hilbert-Einstein gravity, while we use the Chern-Simons action for $f$ mode. Of course
we could have an interaction term as well. As a result one may consider the following parity violating
action for the $f$-$g$ gravity
\bea
&&I[g,f]=\frac{1}{16\pi G}\int d^3x\bigg[ \sqrt{-g} \left( R^g+\frac{2}{\ell^2}\right)
+\sigma\sqrt{-f}\epsilon^{abc}\left(\Gamma^{(f)i}_{aj}\partial_b \Gamma^{(f)j}_{ic}+\frac{2}{3}\Gamma^{(f)i}_{aj}\Gamma^{(f)j}_{bk}\Gamma^{(f)k}_{ci}\right)\cr &&\cr
&&\;\;\;\;\;\;\;\;\;\;\;\;\;\;\;\;\;\;\;\;\;\;\;\;\;\;\;\;\;\;\;\;\;\;\;\;\;\;\;\;\;\;\; - U(g,f) \bigg],
\eea
Since the effective potential we are considering is not symmetric under exchanging $f$ and $g$ the
above model is not equivalent to that when the actions of $g$ and $f$ are exchanged.
For the case we are considering the corresponding equations of motion are given by
\be
G^g_{\mu\nu}-\frac{1}{\ell^2}g_{\mu\nu}=T^g_{\mu\nu},\;\;\;\;\;\;\;C^f_{\mu\nu}=\frac{1}{\sigma} T^f_{\mu\nu},
\ee
where $T^g_{\mu\nu}$ and $T^f_{\mu\nu}$ are given in \eqref{tens}, and $C^f_{\mu\nu}$ is the Cotton tensor
associated with metric $f$ given by
\be
C^f_{\mu\nu}={\epsilon_\mu}^{\alpha\beta} \nabla_\alpha (R^f_{\beta\nu}-\frac{1}{4}R^ff_{\beta\nu}).
\ee
This model admits an AdS solution given by $g_{\mu\nu}=f_{\mu\nu}=g^{AdS}_{\mu\nu}$. It would be interesting to study
fluctuations around this vacuum. It would also be interesting to study other solutions of this model too.

\section*{Acknowledgments}
We would also like to thank  Reza Fareghbal and Amir E. Mosaffa for
useful discussions. M. A. Would also like to thank CERN TH-division for very warm hospitality during the
last stage of the project. This work is supported in part by Iranian TWAS
chapter at ISMO.

%
%
%
%


\end{document}